\documentclass[11pt,a4paper]{article}
\pdfoutput=1
\usepackage{jcappub}
\usepackage{blindtext}
\usepackage{hyperref}
\usepackage{amsmath,amssymb}
 \usepackage{microtype}

\usepackage{graphicx}
\usepackage{bm}
\usepackage{latexsym}
\usepackage{epsfig}
\usepackage{psfrag}
\usepackage{color}
\usepackage[dvipsnames]{xcolor}
\usepackage{subfigure}
\usepackage{graphicx}

\usepackage{amssymb}
\usepackage{amsmath}
\usepackage{bm}
\usepackage{latexsym}
\usepackage{epsfig}
\usepackage{psfrag}
\usepackage{amsmath}
\usepackage[normalem]{ulem}
\usepackage{textcomp}
\usepackage{color}
\usepackage{pstricks}
\usepackage[utf8]{inputenc}
\usepackage{array}
\usepackage{comment}
\usepackage{mathalpha}
\usepackage{tabularray}
\usepackage[most]{tcolorbox}
\makeatletter
\gdef\@fpheader{}
\makeatother

\def\beq{\begin{equation}}
\def\eeq{\end{equation}}
\def\bea{\begin{eqnarray}}
\def\eea{\end{eqnarray}}
\def\be{\begin{equation}}
\def\ee{\end{equation}}

\def\bse{\begin{subequations}}
\def\ese{\end{subequations}}

\def\ee{\eta_{\rm e}}

\def\Mpl{M_{P}}
\def\f{\frac}
\def\l{\left}
\def\r{\right}
\def\d{\mathrm{d}}

\def\fpbh{f_{_{\rm PBH}}}

\def\Tin{T_{\mathrm {in}}}

\def\Tbh{T_{\mathrm {BH}}}
\def\Tev{T_{\mathrm {ev}}}
\def\tin{t_{\rm in}}
\def\tq{t_{\rm q}}
\def\tbh{t_{\rm BH}}
\def\tev{t_{\rm ev}}
\def\tbbn{t_{\rm BBN}}

\def\Min{M_{\mathrm {in}}}

\def\mbh{M_{\mathrm {BH}}}
\def\gammabh{\Gamma_{_{\rm{BH}}}}

\def\hin{H_{\rm in}}

\def\aev{a_\mathrm{ev}}
\def\abbn{a_\mathrm{BBN}}
\def\ain{a_{\rm in}}
\def\abh{a_{\rm BH}}
\def\aq{a_{\rm q}}

\def\rhoin{\rho_{\rm in}}

\def\rbh{\rho_{\mathrm{BH}}}
\def\nbh{n_{\mathrm{BH}}}

\allowdisplaybreaks

\begin{document}
%%%%%%%%%%%%%%%%%%%%%%%%%%%%%%%%%%%%%%%%%%%%%%%%%%%%%%%%%%%%%%%%%%%%%%%%%%%%%%%
\title{Quantum effects on the evaporation of PBHs: contributions to dark matter}

\author{Md Riajul Haque$^{1\,a}$}
\emailAdd{$^a$riaj.0009@gmail.com}
\author{Suvashis Maity$^{1\,b}$,}
\emailAdd{$^b$suvashis@physics.iitm.ac.in}
\author{Debaprasad Maity$^{2\,c}$,}
\emailAdd{$^c$debu@iitg.ac.in}
\author{Yann Mambrini$^{3\,d}$,}
\emailAdd{$^d$yann.mambrini@ijclab.in2p3.fr}

\affiliation{$^1$Centre for Strings, Gravitation and Cosmology,
Department of Physics, Indian Institute of Technology Madras, 
Chennai~600036, India}
\affiliation{$^2$Department of Physics, Indian Institute of Technology Guwahati, Guwahati, Assam, India}
\affiliation{
$^3$ Universit\'e Paris-Saclay, CNRS/IN2P3, IJCLab, 91405 Orsay, France
}

%%%%%%%%%%%%%%%%%%%%%%%%%%%%%%%%%%%%%%%%%%%%%%%%%%%%%%%%%%%%%%%%%%%%%%%%%%%%%%%

\abstract
{We compute the relic abundance of dark matter in the presence of Primordial Black Holes (PBHs) beyond the semiclassical approximation. We take into account the quantum corrections 
due to the memory burden effect, which is assumed to suppress the 
black hole evaporation rate by the inverse power of its own entropy. Such 
quantum effect significantly enhances the lifetime, rendering the 
possibility of PBH mass $\lesssim 10^{9}$ g being the sole dark matter (DM) candidate. 
However, Nature can not rule out the existence of fundamental 
particles such as DM. We, therefore, include the possibility of 
populating the dark sector by the decay of PBHs to those 
fundamental particles, adding the contribution to stable PBH whose lifetime is extended due to the quantum corrections. Depending on 
the strength of the burden effect, we show that a wide range of 
parameter space opens up in the initial PBH mass and fundamental dark matter mass plane that respects the correct relic abundance. 
 } 

\maketitle

%%%%%%%%%%%%%%%%%%%%%%%%%%%%%%%%%%%%%%%%%%%%%%%%%%%%%%%%%%%%%%%%%%%%%%%%%%%%%%%
\section{Introduction \label{sec: Introduction}}
Primordial Black holes (PBHs) are fascinating objects 
proposed more than 50 years ago \cite{Carr:1974nx}, 
but they have recently gained widespread interest, 
particularly in the context of dark matter (DM) \cite{Carr:1974nx,Hawking:1971ei,Ivanov:1994pa,Bartolo:2018evs,Carr:2020xqk,Jedamzik:2020ypm,Jedamzik:2020omx,Green:2020jor,Villanueva-Domingo:2021spv,Carr:2021bzv,Cai:2018dig} and searches
for gravitational waves \cite{Baumann:2007zm,Espinosa:2018eve,Domenech:2019quo,Ragavendra:2020sop,Inomata:2023zup,Franciolini:2023pbf,Firouzjahi:2023lzg,Maity:2024odg}. Indeed, 
contrary to the conventional DM matter candidates, the proposal of PBHs as a possible relic could be considered a huge conceptual 
jump, which does not require any new physics beyond the Standard Model and gravity. 
However, black holes are metastable candidates and emit
fundamental particles through the well-known phenomena 
of Hawking radiation \cite{Hawking:1974rv, Hawking:1975vcx}. 
In vanilla scenarios, PBHs weighing less than $10^{15}$ grams have already evaporated, so they cannot be considered as potential dark matter candidates, even if they can still play an important role in reheating \cite{RiajulHaque:2023cqe}, leptogenesis \cite{Bernal:2022pue,Barman:2024slw,Calabrese:2023key}, gravitational wave 
\cite{Sugiyama:2020roc,Inomata:2020lmk,Domenech:2020ssp,Papanikolaou:2020qtd,Domenech:2021wkk,Papanikolaou:2022chm,Bhaumik:2022pil,Bhaumik:2022zdd,Ghoshal:2023sfa,Balaji:2024hpu,Wang:2022nml} or dark matter production
\cite{Green:1999yh,Cheek:2021odj,Haque:2024cdh,Cheek:2021cfe,Bernal:2022oha}.
In fact, even PBHs of higher masses, in spite of having survived, %the same Hawking effect 
enter in conflict with observations of $\gamma$-ray \cite{Page:1976wx,MacGibbon:1991vc,BALLESTEROS2020135624}, light elements abundance during big bang nucleosynthesis (BBN) \cite{Clark:2016nst,PhysRevD.61.023501,Carr:2009jm}
or other indirect observations \cite{Poulin:2016anj,Niikura:2017zjd,Croon:2020ouk,Griest:2013aaa}. Therefore, the idea of PBHs being a real DM candidate is gradually fading away except within a very narrow mass window $10^{16} - 10^{23}$ grams (see the review \cite{Carr:2020gox} and references therein for a complete analysis). 

However, this semiclassical phenomenon of black hole radiation heavily relies on the assumption that during the 
evaporation process, the black hole remains classical till the end of its lifetime. Over the years, it has been realized that such a semiclassical approach 
should not remain self-similar \cite{Dvali:2012en,Michel:2023ydf,Wang:2023wsm} throughout its entire lifetime, and new physics should come into play, 
particularly in the parlance of the information loss paradox. It has recently been argued \cite{Dvali:2020wft} that during its 
evaporation, when the mass of a BH reaches a certain fraction of its initial value, the backreaction can not 
be ignored, and it can potentially reduce the evaporation rate by the inverse power law of its entropy $S^{-k}$. This, in turn, 
significantly enhances the lifetime of black holes, reopening the possibility of the BHs of mass $<10^{15}$ grams being the sole DM 
candidate as recently discussed in \cite{Alexandre:2024nuo,Thoss:2024hsr}. 
In this paper, we will be inclusive, considering also the possibility of a multi-component scenario, where part of the total DM could be fundamental particles produced through the evaporation process. Such inclusion and its implicit dependence on the enhanced BHs lifetime are, therefore, expected to alter the existing constraints \cite{Carr:2009jm,Carr:2020gox} on particle DM and PBH parameter space significantly, which is the subject of our study.

At this point, we shall mention that there are several potential mechanisms through which PBHs can form. The mechanism includes the collapse of the enhanced density perturbations that originated during  inflation~\cite{Ivanov:1994pa,PhysRevD.50.7173,Yokoyama:1998pt,
Kohri:2007qn,Harada:2013epa,Bhattacharya:2019bvk, Ragavendra:2020sop,Maity:2024odg},
bubble collisions~\cite{Hawking:1982ga}, 
collapse 
of domain walls~\cite{Rubin:2001yw,Dokuchaev:2004kr}, 
collapse of cosmic strings~\cite{HOGAN198487,HAWKING1989237}, 
electroweak phase transition~\cite{Huber:2015znp}, 
first-order phase transitions~\cite{Caprini:2015zlo, 
Dev:2016feu}, 
and formation of other topological defects~\cite{Figueroa:2012kw,Sanidas:2012ee,Dvali:2021byy}.
In this work, we will not focus on the details of how the PBHs are formed. Rather, our discussion spans around the evolution, especially the evaporation of the PBHs and the effects of memory burden on the process.

The paper is organized as follows. In section \ref{sec: pbh evaporation}, we discuss the evolution of PBH mass when the memory effect is active after a certain fraction of the initial mass of the PBH remains due to Hawking evaporation.
In section \ref{sec: dm production}, we analyze the effects of the memory burden on the DM that is produced due to the evaporation of PBHs.
We shall distinguish two different scenarios in this context. In section \ref{sec:dm-full-evap}, we consider the mass range of PBHs that evaporates {\it before} BBN completely, whereas in section \ref{sec:dm-stable-evap}, we look at the case where the initial phase of evaporation happens before BBN while the PBHs are 
{\it stable} till present day. We then conclude in section \ref{Sec:conclusion}.

Before we begin, let us list out the main notations that we will be using throughout our analysis:

\begin{tcolorbox}[colback=gray!5!white,colframe=gray!50!black,colbacktitle=gray!75!black,title=Notations about scale factor and time]
\begin{itemize}
\item [] $\ain,\tin:$ Scale factor and time  at the PBH formation
\item [] $\aq,\tq:$ Scale factor and time at the onset of memory burden 
\item [] $\abh,\tbh:$ Scale factor and time at PBH domination
\item [] $\aev,\tev:$ Scale factor and time at the PBH evaporation 
\item [] $\abbn, \tbbn:$ Scale factor and time at BBN
\end{itemize}
\end{tcolorbox}
\noindent 

%The production of PBHs is usually attributed to the enhanced inflationary scalar power spectrum which directly leads to the enhanced density contrast. Therefore, fraction of PBHs contributing as a DM candidate could      

%%%%%%%%%%%%%%%%%%%%%%%%%%%%%%%%%%%%%%%%%%%%%%%%%%%%%%%%%%%%%%%%%%%%%%%%%%%%%%%
\section{Mass evolution of PBHs due to evaporation an the effects of memory burden\label{sec: pbh evaporation}}
\

As a first step, we need to study the evaporation process of the black hole in detail, taking into account the memory burden effects (for a detailed discussion see Refs. \cite{Dvali:2020wft,Alexandre:2024nuo,Thoss:2024hsr}).
%We need our discussion with the evaporation process of PBHs. 
After the formation, PBH energy density $\rbh=\nbh\mbh$ is mostly governed by two quantities: the PBH number density, $\nbh$ which decreases as the universe expands 
%and with the scale factor $a$, it behaves 
as $\nbh\sim a^{-3}$ ($a$ be the scale factor),
and the mass of PBHs $\mbh$, which changes due to the mechanisms of evaporation and accretion.
We shall first present a brief summary of the evaporation process
in the semiclassical approximation before considering the quantum effect. 
The Hawking temperature, $\Tbh$ and entropy, $S$ associated with a PBH with mass $\mbh $ are given by
\bea 
\Tbh=\f{\Mpl^2}{\mbh},\quad\quad S=\f{1}{2}\l(\f{\mbh}{\Mpl}\r)^2=\f{1}{2}\l(\f{\Mpl}{\Tbh}\r)^2,
\label{eq: temp entropy}
\eea 
where $\Mpl= 1/\sqrt{8 \pi G_N}\simeq 2.4\times 10^{18}$ GeV
is the reduced Planck mass.
If we consider that the evaporation happens by purely Hawking radiation then the rate of change of mass at any point is \cite{Hawking:1974rv}

%a function of the product of the PBH density and the surface area of PBH. It is straightforward to show that the mass rate at a particular time is proportional to the inverse square of PBH mass at that time. The exact expression is given by
\bea 
\f{\d \mbh}{\d t}=-\epsilon \f{\Mpl^4}{\mbh^2},
\label{eq: mass evolution 1}
\eea 
where $\epsilon =\f{27}{4}\f{\pi g_{\ast}(\Tbh)}{480}$, 
with $g_{\ast}(\Tbh)$ being the number of degrees of freedom 
associated with the PBH temperature. The factor ${27}/{4}$ accounts for the graybody factor\footnote{For instance, a more thorough formulation for the greybody factor can be found in Refs.~\cite{Auffinger:2020afu,Masina:2021zpu,Cheek:2021odj}.} \cite{MacGibbon:1990zk}, and the negative sign on the right-hand side is since the PBH mass decreases with time due to the evaporation. We shall note that the accretion effect is negligible here.

After integrating Eq.~\eqref{eq: mass evolution 1} from the time of the formation of PBH, $\tin$ to the time $t$, we find the PBH mass at $t$ is given by
\bea 
\mbh(t)=\Min\l[1-\gammabh^0(t-\tin)\r]^{1/3},
\label{eq: mass 1}
\eea 
where $\Min$ is the initial mass of the PBH, which is related to
the horizon size at the time of formation, 
\beq
\Min=\frac{4}{3} \pi \gamma \rhoin \hin^{-3}
= 4\pi\gamma\Mpl^2/H_{\rm in}\,,
\label{Eq:min}
\eeq
where we assume the formation of PBHs in a radiation-dominated universe, so $\hin^2={\rho_{R}(\ain)}/{(3 \Mpl^2)}$, and $\rho_{R}(\ain)$ is the background radiation energy density and can be connected with the formation temperature $T_{\rm in}$, $\rho_{R}(\ain)=\alpha\,T_{\rm in}^4$ with $\alpha=\pi^2 g_*/30$. Here, $g_*$ is the relativistic degrees of freedom associated with the thermal bath, which we assume $g_*=106.75$. 
In Eq.~\eqref{Eq:min},
 $\gamma$ represents the efficiency factor for collapse, which defines what fraction of the total mass inside the Hubble radius collapses to form PBHs. For standard radiation domination, $\gamma\sim 0.2$ \cite{Carr:1974nx}. 
The time of the PBH formation can be related to the formation mass of the PBH as 
$\tin=\Min/(8\pi\gamma\Mpl^2)$, 
where we have considered a radiation-dominated Universe, $H(t)={1}/{(2t)}$. 
The quantity $\gammabh^0=3\epsilon \Mpl^4/\Min^3$ is the decay width associated with the evaporation of PBH. 
Note that the evolution of the PBH mass
described by Eq.~\eqref{eq: mass 1} is really abrupt, hence Stephen Hawking's own description of it as an `explosion'. The mass of the PBH $\mbh(t)$ remains almost constant 
$\mbh\sim \Min$ during the whole process of evaporation, and then the PBH
rapidly decays at $t=\tev$.

We find the lifetime, $t_{\rm ev}$
of the PBH by solving $\mbh(t_{\rm ev})=0$ in Eq.~\eqref{eq: mass 1} 
\bea 
\tev=\f{1}{\gammabh^0}=\f{\Min^3}{3\epsilon \Mpl^4}
\simeq 2.4\times 10^{-28} \l(\f{\Min}{1~g}\r)^3 s,
\label{eq: tev 1}
\eea 
where we supposed $\tev \gg \tin$.
During radiation domination, the Hubble parameter 
is related to the radiation temperature as $H=\sqrt{\frac{\alpha}{3}}\,\frac{T^2}{\Mpl}$.
One can then connect time and temperature during the radiation era as 
\bea 
t=\f{1}{2}\sqrt{\f{3}{\alpha}}\f{\Mpl}{T^2}\,.
\label{eq: time temp}
\eea 
From Eq.~\eqref{eq: tev 1}, we recover that PBHs of mass $\gtrsim 10^{15}$ grams
should not have decayed yet, whereas PBHs of mass below $\lesssim 10^9$ grams should have decayed before the BBN epoch, or equivalently, before 1 second\footnote{Note that all the above conclusions are under the assumption that the PBHs radiate particles in a self-similar semiclassical process until their end of life.}. 
\begin{figure}
    \centering
    \includegraphics[scale=.37]{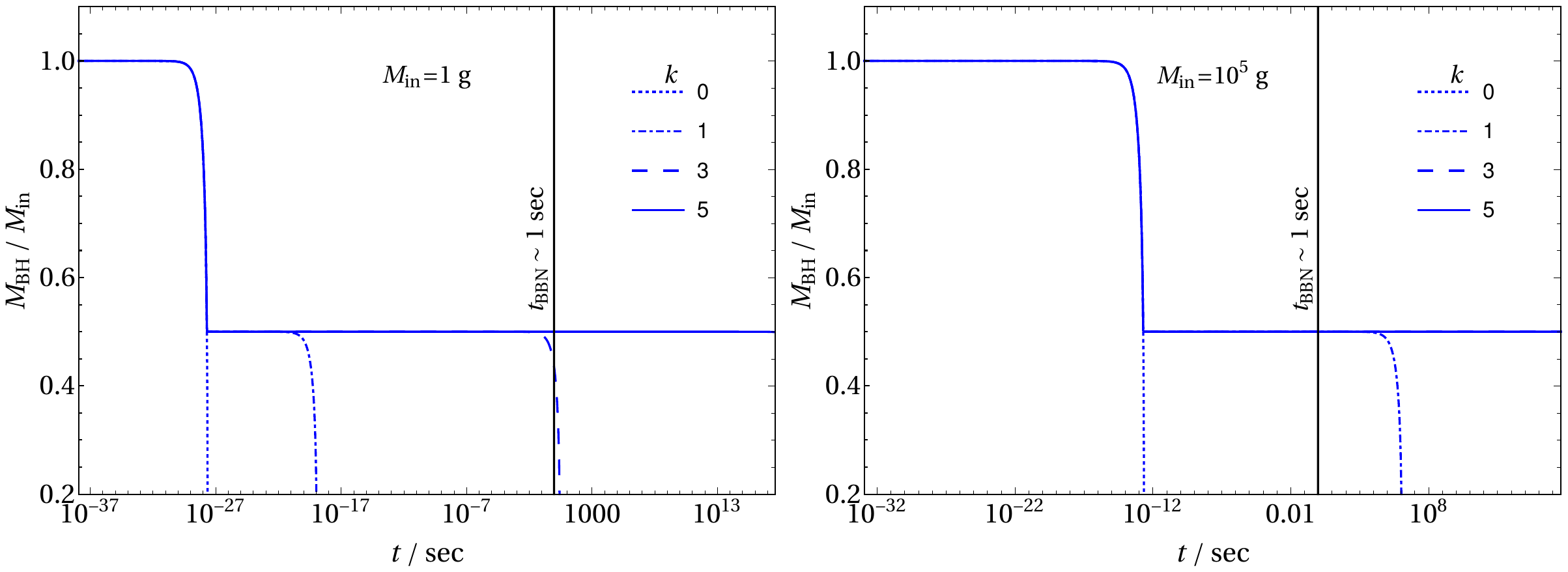}
    \caption{
    \it  Evolution of the mass of the PBH as a function of time for different values of $k$ are plotted here. 
    In the left plot, we have chosen $\Min=1$ g, and in the right plot, we have chosen $\Min=10^5$ g and $q$ is set to the value $q=1/2$. We have chosen four values of $k$, where $k=0$, $1$, $3$, and $5$ are plotted in dotted, dot-dashed, dashed, and solid lines, respectively. We have taken the maximum value of time to be the current age of the universe, $t\sim 4\times 10^{17}$ sec. 
    We see the lifetime of a PBH increases with $k$.
   }
    \label{fig: M-in k}
\end{figure}

We now consider a special type of correction to the evaporation process proposed in \cite{Dvali:2020wft}. Such correction is called the memory burden's effect, which essentially suggests that the quantum modes associated with the entropic degrees of freedom of a black hole necessarily have a strong backreaction effect on its own evaporation process. Therefore, the initial semiclassical Hawking evaporation will no longer be valid after a certain time scale. For the purpose of our analysis, we can suppose that the semiclassical regime is valid
until the mass of the PBH reaches a certain value, 
i.e. $\mbh=q\Min$, with $0<q<1$. 
The authors of \cite{Alexandre:2024nuo,Thoss:2024hsr} 
proposed that the quantum
effects begin to be important when $\mbh=({1}/{2})\Min$,
or $q={1}/{2}$. However, such value is subjected to the detailed quantum mechanical modeling of a black hole. 
To keep our study as general as possible, let $\tq$ be the time at the end of the semiclassical phase.
Then, from Eq.~\eqref{eq: mass 1} we obtain 
\bea 
\tq= \f{1-q^3}{\gammabh^0},
\label{eq: tq}
\eea 
where $\gammabh^0$ is defined before.
In the above equation, with the substitution of $q=0$, one shall recover the full evaporation time, $\tev$.
Once the mass of the PBH reaches $q\Min$, at $\tq$,
the quantum memory effect starts dominating. Upon parameterizing 
the memory burden effect in the second phase, the evolution of the mass which is given in Eq.~\eqref{eq: mass evolution 1} modifies to \footnote{ In Ref.~\cite{Thoss:2024hsr}, the effect of memory burden comes as $S(q M_{\rm in})^{-k}$, where $q M_{\rm in}$ is the value of the PBH mass where the quantum effect starts. Thus, they took a constant correction factor. However, in our work, we have taken the correction through the variable mass-dependent entropy $S( M_{\rm BH})^{-k}$, which is followed by the idea given in \cite{Alexandre:2024nuo}. We have checked that our results are roughly the same in both scenarios since the mass of the PBH remains almost constant throughout the evolution and only starts to dissipate at the very end of evaporation. }
\bea 
\f{\d \mbh}{\d t}=-\f{\epsilon }{\l[S(\mbh)\r]^k}\f{\Mpl^4}{\mbh^2},
\label{eq: mass evolution 2}
\eea 
where $S(\mbh)$ is the BH entropy defined in Eq.~\eqref{eq: temp entropy}. The parameter $k$ characterizes the efficiency of the backreaction effect. So far, we do not have any theoretical constraints on the value of the power $k$ except for its probably being a positive number. For our present purpose, we would take it to be an integer.
We can understand the quantum effect as a slowdown of the decay due to an 
excess of entropy, produced by its evaporation, surrounding the PBH. 
After integrating Eq.~\eqref{eq: mass evolution 2} we obtain
\bea 
\mbh=q\Min\l[1-\gammabh^k(t-\tq)\r]^{\frac{1}{3+2k}},\quad \text{with} \quad
\gammabh^k= {2^k(3+2k)\,\epsilon}\Mpl
\l(\f{\Mpl}{q\Min}\r)^{3+2k}.
\label{eq: mass 2}
\eea 
%%%%%%%%%%%%%%%%%%%%%%%%%%%%%%%%%%%%%%%%%%%%%%%%

%%%%%%%%%%%%%%%%%%%%%%%%%%%%%%%%%%%%%%%%%%%%%%%%
%%%%%%%%%%%%%%%%%%%%%%%%%%%%%%%%%%%%%%%%%%%%%%%%
\begin{figure}
    \centering
    \includegraphics[scale=.57]{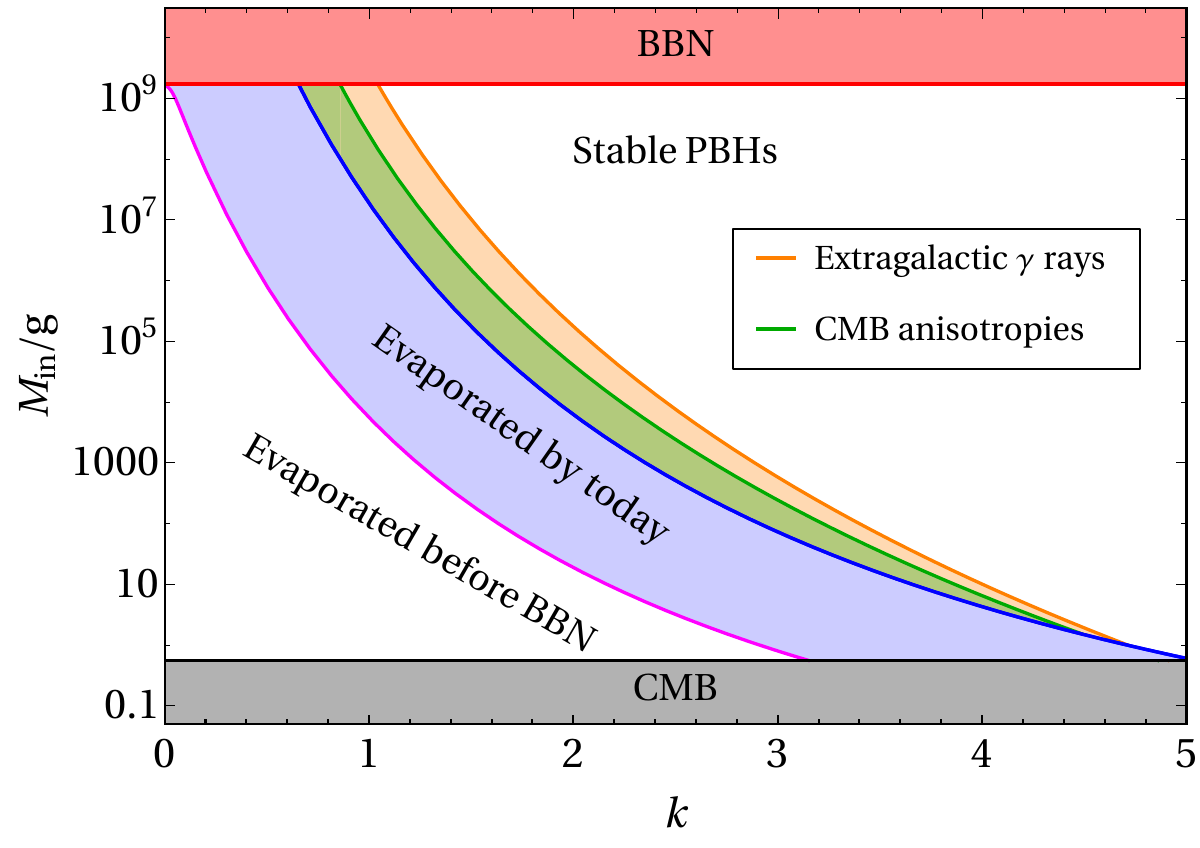}
    \caption{\it  Different limits on the formation mass as a function of $k$ are plotted here. The red line represents the $\Min$ above which the half-life of PBH will be after BBN.  The region above the blue line represents $\Min$ for the stable PBHs today. The region below the magenta line represents $\Min$ for the PBHs that evaporated before BBN. The black line corresponds to the minimum allowed value of the formation mass, $\Min$ calculated from the maximum allowed value of the inflationary energy scale $H_{\rm inf}\sim 5\times 10^{13}$ GeV, set by the upper bound of the tensor-to scalar ratio $r<0.036$ from CMB data \cite{BICEP:2021xfz}. 
    The orange and the green shaded regions are the constraints from the Extragalactic $\gamma$ rays and CMB anisotropies, respectively, on sable PBHs where PBHs cannot be total dark matter relic observed today, which indicates $\fpbh<1$,  $\fpbh$ being the ratio between the PBH and the dark matter energy density calculated today (see the text for details). For the restriction from Extragalactic $\gamma$ rays and CMB anisotropies, we use Ref. \cite{Thoss:2024hsr}.}
    \label{fig: Mpbh k}
\end{figure}
%%%%%%%%%%%%%%%%%%%%%%%%%%%%%%%%%%%%%%%%%%%%%%%%

From Eq.~\eqref{eq: mass 2}, we obtain that the second phase of evaporation will occur for a time $\sim 1/\gammabh^k$. Hence, the total evaporation
time is given by $\tev^k=\tq+1/\gammabh^k\simeq 1/\gammabh^k$. 
It is straightforward to see that for $k=0$ and $q=1$, $\tev^{k}$ will be equal to as given in Eq.~\eqref{eq: tev 1}. Also for $k>0$, we have $\tev^{k}\gg \tev$ (for a detailed discussion see Ref.~\cite{Alexandre:2024nuo}) and
we obtain

\beq
\tev^k \simeq \frac{q^{3+2k}}{2^k(3+2k)}\left(\frac{\Min}{4.3\times10^{-6}~\rm{g}}\right)^{3+2k}~5.7 \times 10^{-44}~\rm{s\,,}
\eeq
which gives, for instance, for $k=1$ and $q=1/2$

\beq
t_{\rm ev}^{k=1} \simeq  ~1.2\times 10^{-19}\left(\frac{\Min}{1~\rm{g}}\right)^5~{\rm s}
\simeq
 ~4\times 10^{17}\left(\frac{\Min}{2\times10^7\,\rm{g}}\right)^5~\rm{s}\,.
\eeq
We then conclude that even in the minimal $k=1$ case, 
a PBH of only $2\times 10^7$ grams can survive until the present time and contribute to the DM relic abundance.
To illustrate our result, we show in the left panel of Fig.~\ref{fig: M-in k}  the evolution of PBH mass as 
a function of time for $q=1/2$ and different values of 
$k=0,\,1,\,3$ and $5$, which are plotted in dotted, dot-dashed, dashed and solid lines respectively. Obviously, we see that for $k=0$, there is no second phase of evolution because the treatment is completely semiclassical.
As the value of $k$ increases, the greater is the quantum effect, slowing down the process of decay even further.
We also presents some numerical results of
the evaporation time $\tev$ in Table~\ref{tb: 1} for 3 PBHs masses (1, $10^4$ 
and $10^8$ g) with and without taking into account the memory burden effect for different values of $k$, and with the 
formation time $\tin$ for information.
%are the second phase of evolution happens more slowly hence the evaporation time increases.

\begin{table}[t!]
  \begin{center}
    \begin{tblr}{|c|c|c|c|c|c|}
      \hline
         $\Min$ (g)& $\tin$ (sec) & $\tev^{k=0}$ (sec) & $\tev^{k=1}$ (sec)& $\tev^{k=3} $ (sec) & $\tev^{k=5} $ (sec) \\ \hline
         $1$& $1.24 \times 10^{-38}$ & $2.34\times 10^{-28}$ & $1.95\times 10^{-19}$ &$8.63$ & $3.82\times 10^{20}$\\ \hline
         $10^4$& $1.24 \times 10^{-34}$ & $2.34\times 10^{-16}$ & $19.489$ &$8.63\times 10^{36}$ & $3.82\times 10^{72}$\\ \hline
         $10^8$& $1.24 \times 10^{-30}$ & $2.3\times 10^{-4}$ & $1.94\times 10^{21}$ &$8.63\times 10^{72}$ & $3.82\times 10^{124}$\\ \hline
    \end{tblr}
    \caption{\label{tb: 1}{
    The time of formation and evaporation for different $\Min$ are tabulated here. $k=0$ represents the case of pure Hawking radiation.
    }}
  \end{center}
\end{table}

We focus on the case where the semiclassical phase of the evaporation ends before BBN. The maximum initial 
mass of a PBH for which it will half evaporate before BBN can be obtained 
from Eq.~\eqref{eq: tq} with $q=1/2$. This is independent of the value of $k$ 
and is represented by the red line on top of the right panel in  Fig.~\ref{fig: M-in k}. The red-shaded region above the line indicates the region where the Hawking evaporation of PBHs happens after BBN, i.e., $\tq>\tbbn$. Note that we assume $\tbbn=1$ second throughout our analysis. On the other hand,
the magenta line corresponds to the maximum value of $\Min$ for which the complete evaporation happens before BBN, i.e., $\tev^{\mathrm{tot}}\leq \tbbn$ and the value is given by
\bea 
\Min=\f{\Mpl}{q} \l({(3+2k)\,\epsilon\, 2^k\, \tbbn \Mpl}\r)^{1/(2k+3)}\,.
\eea 
We see that with the increase in the value of $k$, the PBHs become more stable; hence, the maximum value of $\Min$ evaporates before BBN decreases.

However, the interesting situation is when PBHs' lifetime in the semiclassical approximation is {\it below} the BBN limit (below the red line) while the quantum effect renders them stable at the scale of the age of the Universe. This region appears to be above the blue line in  Fig.~\ref{fig: M-in k}.
In other words, considering the quantum effect, some of the PBHs with mass $\Min\leq 10^9$ g will survive till today. For instance, for $k=1,~2$, and $3$, the lower limit of the mass that should have survived are 
 $\sim 2\times 10^7$ g, $7\times 10^3$ g, and $80$ g respectively.
However, being stable at the scale of the age of the Universe is not sufficient. 
Indeed, the metastability of PBHs implies the possibility to observe them through 
their radiation in the CMB or extragalactic $\gamma-$ray. The non-
observation of such signals exclude part of the region, shaded in 
green and orange of Fig.~\ref{fig: Mpbh k} respectively, to be the 
sole DM candidate. Therefore, if PBHs form in those mass ranges at all, their fractional abundance must be very 
small, and an additional component needs to be incorporated to match the present DM abundance under the form of dark particles, for instance. We will calculate 
in detail those contributions of additional particles like DM 
components produced directly from the PBHs decay.

To be more precise, PBHs can act as DM due to their cold, non-interacting nature.
The fraction of total DM that is given by the stable PBHs today is often referred to as $\fpbh$.
In Fig.~\ref{fig: Mpbh k}, the region above the blue line has two additional constraints on $\fpbh$ coming from extragalactic $\gamma$ rays and from CMB anisotropies which are shaded in orange and green, respectively. PBHs can not contribute to the total dark matter in these regions, i.e., $\fpbh<1$.  
The contribution to the extragalactic $\gamma$ rays comes from the evaporation of PBHs between the time of recombination and the present time.
On the other hand, the evaporation of PBHs can inject energy into the neutral medium and ionize it after the recombination.
This restricts the amount of $\fpbh$ as the injected energy affects the angular power spectrum of temperature and polarization of CMB due to the rescattering of CMB photons. A detailed discussion about these constraints can be found in Ref. \cite{Thoss:2024hsr}.

As emphasized earlier, in this paper, we consider the present-day DM to be composed of both PBHs and fundamental particles. Our work will then focus on these two interesting regions, the unshaded ones in Fig.~\ref{fig: Mpbh k}, 
where dark matter can be produced by PBH evaporation while not being excluded by other constraints.
Indeed, whereas PBHs can be dark matter candidates
within this parameter space, the fact that the semiclassical
process takes place before BBN also creates the possibility 
of producing a particle-like dark matter candidate through the PBH's early decay. In that case, the relic abundance should be the sum of the PBH dark decay product and the surviving PBH density.
Note that the black line corresponds to the minimum value of $\Min$ set by the maximum allowed value of the inflationary energy scale $H_{\rm inf}\sim 5\times 10^{13}$ GeV, calculated considering the upper bound of the tensor-to scalar ratio $r<0.036$ from Planck together with latest BICEP/$Keck$ data \cite{Planck:2018jri,BICEP:2021xfz}.

%%%%%%%%%%%%%%%%%%%%%%%%%%%%%%%%%%%%%%%%%%%%%%%%%%%%%%%%%%%%%%%%%%%%%%%%%%%%%%%
\section{Production of dark matter and the effects of memory burden \label{sec: dm production}}

To study the effects of the memory burden on the production of dark matter from the evaporation of PBHs, we should consider two possible scenarios. 
We shall first consider the case when the PBHs evaporate completely before BBN, and the dark matter generated from the evaporation satisfies the present dark matter relic. 
This is represented by the white region below the magenta line in Fig.~\ref{fig: Mpbh k}. 
Another possibility is to consider that the first phase of Hawking evaporation (semiclassical approximation) 
happens {\it before} the BBN, whereas the PBH remains 
stable till today due to the memory effect.  
This is the white region above the blue line of Fig.~\ref{fig: Mpbh k}. 
First, we need to compute the number of particles that are emitted from the evaporation of one PBH. 
Note that from now on, we will coin {\it phase-I}  the semiclassical evaporation phase,  and {\it phase-II} the second phase, where quantum correction is effective.

%%%%%%%%%%%%%%%%%%%%%%%%%%%%%%%%%%%%%%%%%%%%%%%%%%%
\subsection{Number of dark matter particles emitted from the evaporation of a PBH}

Let $N_j$ be the number of particles that are emitted from the evaporation of one PBH. 
We divide $N_j$ into two parts $N_{1j}$ and $N_{2j}$, which are the number of particles emitted from PBHs for {\it phase-I} and {\it phase-II}, respectively.
We shall mention that BH mass and spin affect the production rate of any species that is generated from evaporation. 
In this work, for simplicity, we restrict ourselves to the case of spin-zero
Schwarzschild BH.
The emission rate of particles $j$ with mass $m_j$ and internal degrees of freedom $g_j$ that escape the horizon of radius $R_S$ 
per unit time per unit energy interval due to the Hawking radiation in the {\it phase-I} is
given by \cite{Haque:2024cdh,RiajulHaque:2023cqe}

\bea 
\f{\d^2 N_{1j}}{\d E\d t} &=& \f{27}{4}\pi R_S^2 \f{g_j}{2\pi^2} \f{E^2}{\exp(E/\Tbh)\pm 1} = \f{27}{4} \f{g_j}{32 \pi^3}\f{(E/\Tbh)^2}{\exp(E/\Tbh)\pm 1}
\label{eq: emission rate 1}
\eea 
where the Schwarzchild radius is given by 
$R_S=\mbh/(4\pi\Mpl^2)$. 
The sign $\pm$ is used for fermionic and bosonic particles, respectively.
We can estimate the total energy emitted per unit time from a BH by integrating over Eq.~\eqref{eq: emission rate 1} as
\bea 
\f{\d N_{1j}}{\d t} =\f{27}{4} \f{\xi g_j  \zeta(3)}{16\pi^3} \f{\Mpl^2}{\mbh}, \quad \text{where}
\quad \xi=\begin{cases}
    1& \text{for bosons}\\
    \f{3}{4}& \text{for fermions}
\end{cases}.
\label{eq: emission 1}
\eea 
In the {\it phase-II},
the emission rate is reduced, affected by the quantum corrections
\bea 
\f{\d^2  N_{2j}}{\d E\d t} = \f{1}{\l[S(\mbh)\r]^k}\f{\d^2 N_{1j}}{\d E\d t} \,,
\eea 
where the expression for $\d^2N_{1j}/(\d E\d t)$ is given in Eq.~\eqref{eq: emission rate 1}. After integrating for all the energy modes, the emission rate is given by 
\bea 
\f{\d N_{2j}}{\d t} =\f{27}{4} \f{\xi g_j  \zeta(3)2^k}{16\pi^3} \f{\Mpl^{2+2k}}{\mbh^{1+2k}}.
\label{eq: emission 2}
\eea

The total number of particles emitted from the evaporation of a BH can be obtained by integrating Eq.~\eqref{eq: emission 1} and Eq.~\eqref{eq: emission 2}. 
Depending on the mass of the emitted particle 
relative to the PBH formation temperature $\Tbh^{\rm in}$, 
one should distinguish two cases. 
If $m_j<\Tbh^{\rm in}$, the production of 
particles happen throughout the lifetime of the PBH, i.e., from the initial time $\tin$ to a final time $\tev$. Indeed, the BH
temperature {\it increasing} with time, the relation 
$m_j< \Tbh$ will always hold. On the other hand,
if $m_j>\Tbh^{\rm in}$, the evaporation will start from a time $t_j$ where $m_j=\Tbh(t_j)$, the production being exponentially 
suppressed before. This can happen either during {\it phase-I}  ($t_j<\tq$) or during {\it phase-II} ($t_j>\tq$). Combining Eq.~\eqref{eq: mass 1} and Eq.~\eqref{eq: mass 2} we 
can calculate the value of $t_j$ 
\bea 
t_j=
\begin{cases}
{\tev^0}\l[1-\f{\Mpl^6}{\Min^3 m_j^3} \r]& \text{for}~t_j<\tq\\
&\\
\tev^k\l[1-\l(\f{\Mpl^2}{q\Min m_j}\r)^{3+2k} \r]& \text{for}~t_j>\tq\,,
\end{cases}
\label{eq:ti}
\eea  
where $\tq$ is given by Eq.~\eqref{eq: tq} and we supposed $t_j \gg \tin$.

With the relations in hand let us proceed to calculate the number of particle emitted from the evaporation. First we shall consider the case when $m_j<T_{_{\rm BH}}^{\rm in}$. To calculate the number in the first phase, we integrate Eq.~\eqref{eq: emission 1} within the limit $[t_{\rm in},t_{\rm q}]$ which leads to
\bea 
N_{1j}&=& \f{15 \xi g_j\zeta(3)(1-q^2)}{g_{\ast}(\Tbh)\pi^4}\f{\Min^2}{\Mpl^2} ,
\label{eq:N1j 1}
\eea 
where we see the case of complete evaporation due to only Hawking radiation can be obtained by substituting $q=0$. During the second phase of evaporation, we shall integrate Eq.~\eqref{eq: emission 2} from $\tq$ to a final time $t$, and the number of dark matter emitted is given by

\bea 
N_{2j} &=& \f{15 \xi g_j\zeta(3)}{g_{\ast}(\Tbh)\pi^4}\f{q^2\Min^2}{\Mpl^2} \l[1-\l(1-\f{t-\tq}{t_{\rm ev}}\r)^{\f{2}{3+2k}}\r]\,.
\label{eq:N2j 1}
\eea 
The total number of dark matter particles from the evaporation of a PBH can be obtained by adding $N_{1j}$ and $N_{2j}$ 
from Eq.~\eqref{eq:N1j 1} and Eq.~\eqref{eq:N2j 1}. For those BHs evaporated by today, $t=\tev\gg \tq$ and we have the total numbers of DM particles emitted from a single BH
\bea 
N_j=N_{1j}+N_{2j}=\f{15 \xi g_j\zeta(3)}{g_{\ast}(\Tbh)\pi^4}\f{\Min^2}{\Mpl^2}\,,
\eea 
which is the same as the total number of particles emitted from the Hawking evaporation. This is not surprising as the total number of emitted particles due to the evaporation of a PBH does not depend on the evaporation process.

%%%%%%%%%%%%%%%%%%%%%%
For the case $m_j>\Tbh^{\rm in}$,
we have two different scenarios: $t_j<t_{\rm q}$ and $t_j>\tq$. For $t_{j}<\tq$, we get the number of particles emitted 
during {\it phase-I} by simply integrating Eq.~\eqref{eq: emission 1} in the interval $[t_j,\tq]$, no particles being produced before due to the Boltzmann suppression factor,
%where we have used $t_j$ for the case of $t_j<t_{\rm q}$ in Eq.~\eqref{eq:ti} and one can find
\bea 
N_{1j}= \f{15 \xi g_j\zeta(3)}{g_{\ast}(\Tbh)\pi^4} 
 \l(\f{\Mpl^2}{m_j^2} -\f{q^2\Min^2}{\Mpl^2}\r).
 \label{Eq:njphase1more}
\eea 
Where we used $\mbh(t_j)={\Mpl^2}/{m_j}$. 
The number of particles emitted  $N_{2j}$ in the {\it phase-II} 
would be the same as obtained in Eq.~\eqref{eq:N2j 1}. 
On the other hand, if $t_j>\tq$, corresponding to $m_j>\frac{\Mpl^2}{q\Min}$, 
there is obviously no dark matter emission during {\it phase-I}, i.e., $N_{1j}=0$.
The number of particles emitted during the second phase is given by 
\bea 
N_{2j}=\f{15\xi g_j \zeta(3)}{g_{\ast}(\Tbh)\pi^4}\f{q^2\Min^2}{\Mpl^2} 
\l[\l(1-\f{t_j-\tq}{\tev}\r)^{\f{2}{3+2k}} 
-\l(1-\f{t-\tq}{\tev}\r)^{\f{2}{3+2k}}\r],
\eea
where we have integrated Eq.~\eqref{eq: emission 2} in the time interval $[t_j,t]$ with $t_j$ given in Eq.~\eqref{eq:ti} for $t_j>\tq$.
In both cases after complete evaporation, we get the total number of DM particles emitted for
$m_j>\Tbh^{\rm in}$ by setting $t=\tev> t_j\gg \tq$, to be
\bea 
N_j=N_{1j}+N_{2j}=\f{15\xi g_j \zeta(3)}{g_{\ast}(\Tbh)\pi^4}\f{\Mpl^2}{m_j^2},
\label{Eq:mjgTbh}
\eea 
which is again the number of particles 
emitted by pure Hawking evaporation for $m_j>\Tbh^{\rm in}$
\cite{Haque:2024cdh,RiajulHaque:2023cqe,Barman:2024slw}.
In the next 
section, we shall discuss the case where the PBH evaporates completely,
before studying the case where relic PBHs could contribute to the dark abundance, in addition to its decay products.
%%%%%%%%%%%%%%%%%%%%%%%%%%%%%%%%%%%%%%%%%%%%%%%%%%%
\subsection{Dark matter from the evaporation of PBHs before BBN \label{sec:dm-full-evap}}

\subsubsection{Computation of $\beta_{\rm c}$}
\begin{figure}
    \centering
    \includegraphics[scale=.57]{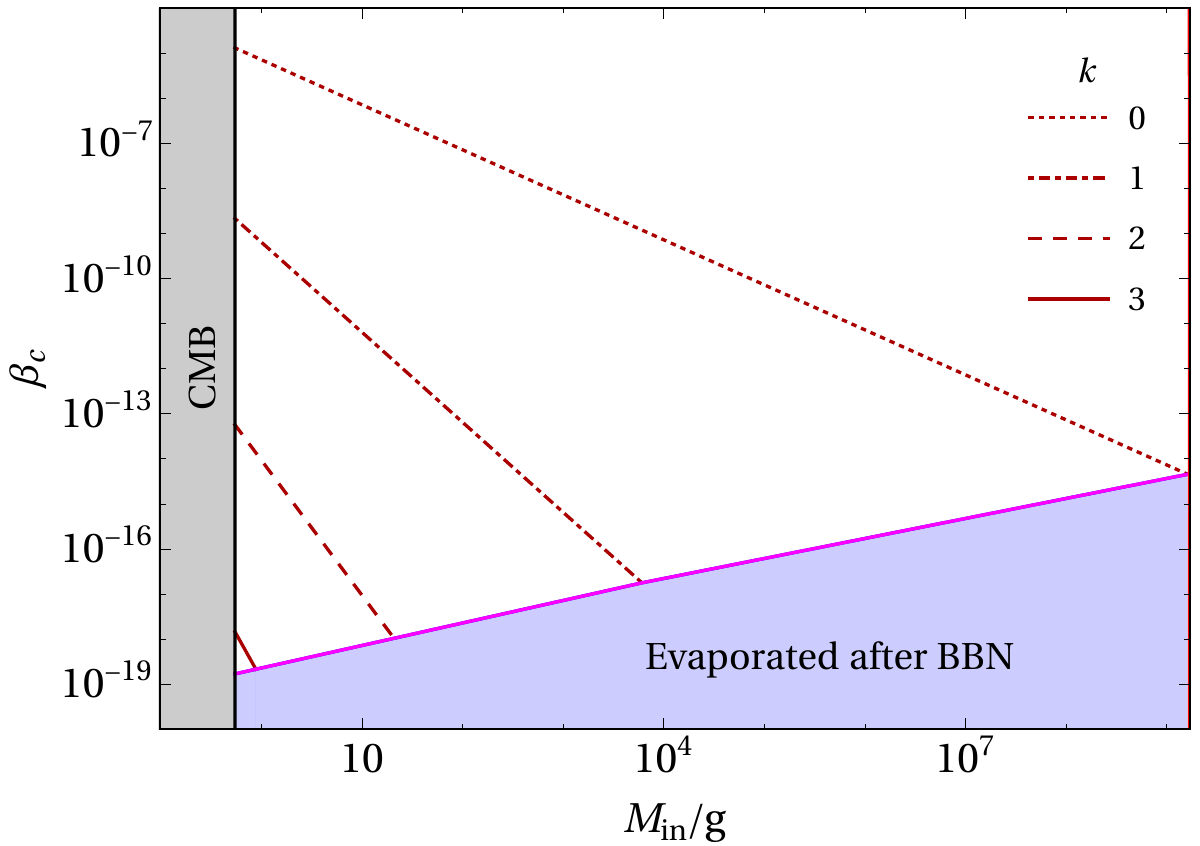}
    \caption{\it Critical values of beta corresponding to the PBH domination are plotted as a function of $\Min$ for different $k$. We have chosen four values of $k$ where $k=0$, $1$, $2$, and $3$ are plotted in dotted, dot-dashed, dashed, and solid lines, respectively. The grey-shaded region is excluded from the minimum PBH mass possible, corresponding to the highest energy scale of inflation. The blue-shaded region corresponds to the PBHs that evaporate after BBN.
   }
    \label{fig:beta-crit}
\end{figure}
The memory burden effect increases the lifetime of the PBHs. Moreover, increasing 
the value of $k$, such burden effect becomes stronger, and as a consequence, it limits the maximum value of 
the PBH mass that evaporates before BBN, 
which is also evident from Fig.~\ref{fig: Mpbh k}. Note that in this figure, we assume 
that the memory effect is effective after its half-life ($q=1/2$). 
On the other hand, the minimum allowed mass
for a PBH is set by the minimum
horizon size at the end of inflation, and determined by the energy available of the inflaton. Assuming de-sitter-like inflation and taking the constraint on the 
tensor-to-scale ratio $r<0.036$ from Planck combined with the latest BICEP/{\it Keck} data \cite{BICEP:2021xfz}, the minimum allowed mass is roughly around $\mbh \gtrsim 0.5$ g. Interestingly, for this minimum PBH mass, one requires $k\lesssim3$ to ensure it is evaporated completely before BBN.

In this section, we will consider two 
possible scenarios for dark matter production, depending on the PBH parameters 
($\Min,\,\beta$), where $\beta = {\rho_{\rm BH}^{\rm in}}/{\rho_R^{\rm in}}$ is the ratio between the PBH energy density over radiation energy density at the point of formation. Initially, we will assume that the radiation energy density always dominates the background. Next, 
we will suppose that for values of $\beta$ above a critical value $\beta_{\rm c}$, 
the PBH energy density dominates over the radiation energy 
density at some time before the evaporation. 
In both cases, we shall assume that the dark matter is produced solely from the evaporation of PBHs. Let's first determine 
$\beta_{\rm c}$.

The radiation energy density evolves as 
\bea \label{Eq: rad}
\rho_{R}(a) &=& \rho_{R}^{\mathrm{in}}\l(\f{\ain}{a}\r)^4,
\eea 
where $\rho_{R}^{\mathrm{in}}$ is the radiation energy density at the time of PBH formation, and $\ain$ is the corresponding scale factor.
We also know that PBHs behave like matter; therefore, the energy density of PBHs varies as $\sim a^{-3}$. At any time, the energy density of PBH can be written in terms of the initial energy density as 
\bea 
\rho_{{\rm BH}}(a)= 
\begin{cases}
    \rho_{\rm BH}^{\rm in}\l(\f{\ain}{a}\r)^3 & \text{for }~a< \aq\\
    q\,\rho_{_{\rm BH}}^{\rm in}\l(\f{\ain}{a}\r)^3 & \text{for }~a> \aq
\end{cases},
\label{eq:rhobh}
\eea
where for $a\sim \aq$, we assume that the mass of the PBHs will be reduced to $q\Min$ instantly. There 
always exists a $\beta$ value, i.e., $\beta_{\rm c}$, above which it is possible 
that after a time $\tbh$, the PBH 
energy density dominates over the radiation energy density.
Equating Eqs.~\eqref{eq:rhobh} and \eqref{Eq: rad}, this happens for a scale factor $\abh$

\beq
\frac{\abh}{\ain}=\frac{1}{q \beta}\,.
\eeq
The domination of PBHs is possible if $\abh<\aev$,
or
\beq
\beta>\frac{\ain}{q\aev}= \frac{1}{q}\sqrt{\frac{H_{\rm ev}}{H_{\rm in}}} = \frac{1}{q}\sqrt{\frac{\gammabh^k}{2 H_{\rm in}}} = \beta_{\rm c}\,,
\eeq
or 
\bea 
\beta_{\rm c}=\frac{1}{q^{\frac52+k}} \l(\f{\Mpl}{\Min}\r)^{1+k} \sqrt{\f{(3+2k)2^k\epsilon }{8\,\pi\gamma}}.
\eea
where we used Eqs.~\eqref{Eq:min} and \eqref{eq: mass 2}.
If one assume $q=1/2$, we have
\bea
\beta_{\rm c}\simeq 5.5\times 2^{\f{3k}{2}}\,\sqrt{\left(3+2k\right)}\left(\frac{\Mpl}{\Min}\right)^{1+k}\,,
\eea
or
\bea
\beta_{\rm c}^{k=0}\simeq 4.0\times 10^{-5}\left(\frac{1~\rm g}{\Min}\right)\,,~~~
\beta_{\rm c}^{k=1}\simeq 6.5\times 10^{-10}\left(\frac{1~\rm g}{\Min}\right)^2\,,~~~
\beta_{\rm c}^{k=2}\simeq 9.4\times 10^{-15}\left(\frac{1~\rm g}{\Min}\right)^3\,.
\nonumber
\eea
As we can notice, the critical value of $\beta$, for a given $\Min$,
decreases drastically with $k$. This is understandable because 
larger $k$ means a longer lifetime. In other words, smaller values 
of $\beta$ are necessary to ensure PBH domination.
%which implies a fixed value of $\Min=1$ g in the scenario of standard Hawking radiation, i.e., $k=0$, $\beta_{\rm c}\sim 10^{-5}$.
As we increase $k$, the $\beta_{\rm c}$ shifted towards smaller values, such as for $k=0$ and $\Min=1$ g, $\beta_{\rm c}\sim 4\times 10^{-5}$,whereas for $k=1$, $\beta_{\rm c}\sim 6\times 10^{-10}$.
We show in Fig.~\ref{fig:beta-crit} the evolution of $\beta_{\rm c}$
as function of $\Min$ for different choices of $k$. Note that even for very light PBHs $\sim 10$ g, a value of $\beta$ as low as $10^{-17}$ is sufficient for the black holes to dominate the evolution of the Universe before BBN for $k=2$.

%%%%%%%%%%%%%%%%%%%%%%%%%%%%%%%%%%%%%%%%%%%%%%%%

%%%%%%%%%%%%%%%%%%%%%%%%%%%%%%%%%%%%%%%%%%%%%%%%
%Now let us calculate the DM abundance from evaporating PBHs for $\beta>\beta_{\rm c}$ and $\beta<\beta_{\rm c}$.

To find the relic abundance of dark matter of the species $j$
today we use \cite{book}, 
\bea
\Omega_jh^2= 1.6\times 10^8\,\frac{g_0}{g_{\rm ev}}\frac{ n_{ j}(\aev)}{\Tev^3}\,\frac{m_j}{\text{GeV}}\,,
\label{Eq:omegah2}  
\eea
where $n_{j}(\aev)$ is the number density of DM at the end of evaporation, whereas $g_{\rm ev}\sim106.75$ and $g_0=3.91$ are the effective numbers of light species for entropy at the end of evaporation and present-day, respectively. 
%We took both effective numbers of degrees of freedom for entropy and radiation is the same.
Once the PBH evaporated, we can write $n_j(\aev)=N_j\,\nbh(\aev)$, where $N_j$ be the total number of particles emitted from the evaporation of a single PBH and $\nbh(\aev)$ is the PBH energy density at the point of evaporation. Let us first discuss the case  $\beta>\beta_{\rm c}$.
%%%%%%%%%%%%%%%%%%%%%%%%%%%%%%%%%%%%%%%%%%%%%%%%%%%%%%%

\subsubsection{$\beta>\beta_{\rm c}$}

\begin{figure}
    \centering
    \includegraphics[scale=.37]{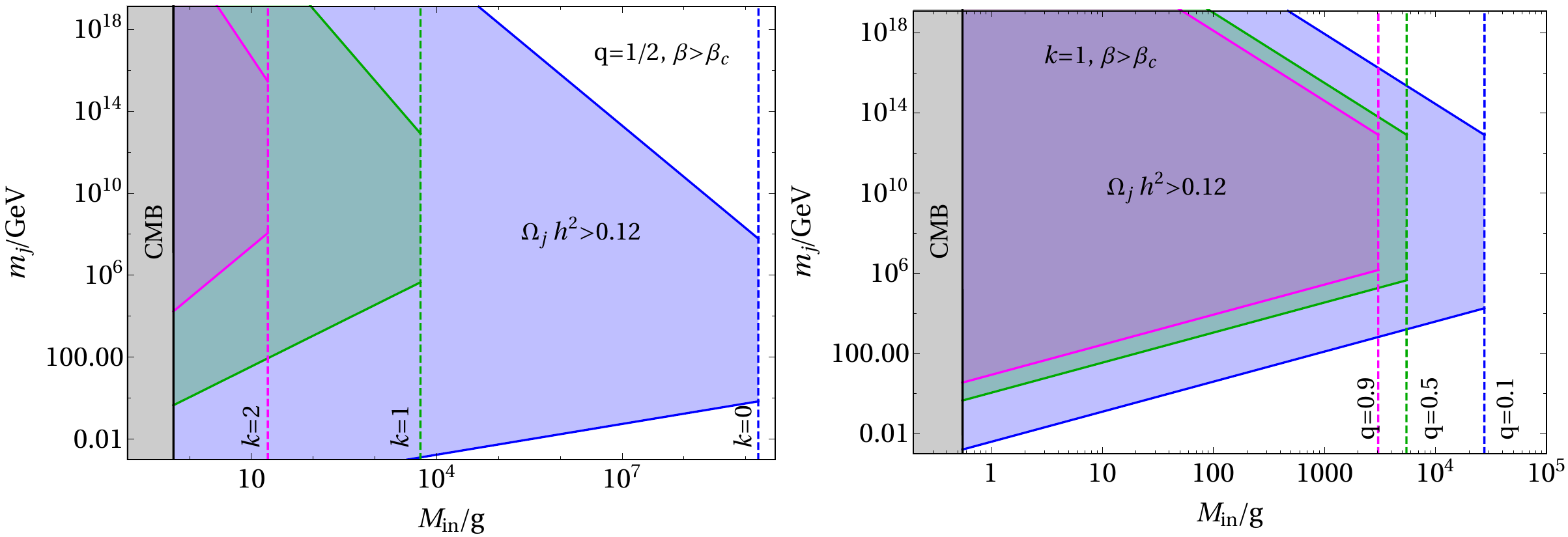}
    \caption{\it The value of the dark matter mass is plotted here as a function of the formation mass of PBHs for the case where the evaporation happens during PBH domination, i.e., $\beta>\beta_{\rm c}$. The black lines represent the minimum mass possible for PBHs calculated considering the maximum energy scale of inflation.
    The shaded regions correspond to dark matter overproduction, $\Omega_j\,h^2>0.12$.
    The vertical dashed lines represent the maximum $\Min$ values allowed to be consistent with the BBN bound.
    {\bf Left panel:} We have chosen $q=1/2$ and plotted for three different values of $k=0,\,1,\,$ and $2$, shown in blue, green, and magenta, respectively.
    {\bf Right panel:} We have chosen $k=0$ and plotted for three different values of $q=0.1,\,0.5,\,$ and $0.9$, illustrated in blue, green, and magenta, respectively.
     }
    \label{fig:mj-min-f1}
\end{figure}
For $\beta>\beta_{\rm c}$, there will be a PBH domination {\it before} PBH decays, which means that the reheating
would then be produced entirely through 
the evaporation. In this case, if one supposes an instant thermalization, the reheating temperature would be the temperature of the decay products given by
\bea
&&
H^2(\aev)=\frac{\rho_{R}(\aev)}{3M_P^2}
=\frac{ \alpha \Tev^4}{3 \Mpl^2}
=\frac{4 \,\left({\gammabh^k}\right)^2}{9}\,,
\label{Eq: beta>betac}
\\
&&
\Rightarrow \Tev=\Mpl\l(\f{4}{3\alpha}\r)^{1/4}
\l[(3+2k)2^k\epsilon\l(\f{\Mpl}{q\,\Min}\r)^{3+2k}\r]^{1/2}\,.
\label{Eq:trh1}
\eea
% with $\alpha={g_{\rm ev} \pi^2}/{30}$.
Since PBH decay happens in a matter-dominated universe,
$H(\tev)={2}/{(3\,\tev)}$. 

From Eq.~\eqref{Eq: beta>betac}, using
$\rho_{\rm BH}(\aev)=\rho_R(\aev)\simeq \nbh(\aev)\times q\Min$, we have the PBH number density at the evaporation point
\bea
\nbh (\aev)=\f{4}{3} \Mpl^3 (3+2k)^22^{2k}\epsilon^{2} \l(\f{\Mpl}{q\,\Min}\r)^{7+4k}\,.
\label{Eq:DMPBH}
\eea
Combining Eq.~\eqref{Eq:omegah2} with Eqs.~\eqref{Eq:trh1} and \eqref{Eq:DMPBH}, writing $n_{j}=\nbh\times N_j$,
we obtain for $m_j<\Tbh^{\rm in}$
\bea
\f{\Omega_jh^2}{0.12}= 2.85\times10^6\,\frac{\xi\,g_j}{q^2}\,\l(2^k(3+2k)\r)^{1/2} \l(\f{\Mpl}{q\,\Min}\r)^{\frac{2k+1}{2}} 
\,\frac{m_j}{\text{GeV}}\,.
\label{Eq:omegah2 3}  
\eea
Doing the same exercise for $m_j>\Tbh^{\rm in}$, Eq.~\eqref{Eq:mjgTbh} gives
\bea
\f{\Omega_jh^2}{0.12}= 1.64\times10^{43}\,{\xi\,g_j}\,\l(2^k(3+2k)\r)^{1/2} \l(\f{\Mpl}{q\,\Min}\r)^{\frac{2k+5}{2}} 
\,\frac{\text{GeV}}{m_j}\,.
\label{Eq:omegah2 4}  
\eea

We show in Fig.~\ref{fig:mj-min-f1} the exclusion parameter 
due to an overabundance of dark matter in the shaded regions, 
for different values of $k$ and $q=1/2$ (on the left)
or $k=1$ and different values of $q$ (on the right). 
We easily recognize the two 
allowed regions, for light masses, corresponding to $m_j\lesssim \Tbh^{\rm in}$, Eq.~\eqref{Eq:omegah2 3}, and for large mass, when the time of allowed decay (when $\Tbh$ reaches $m_j$)
is sufficiently short to avoid the overproduction of dark matter, 
Eq.~\eqref{Eq:omegah2 4}. The allowed region widens for larger values of $k$. This is easily understandable by the fact that 
the memory burden effect extends the lifetime of PBH, diluting them further and then decaying. More diluted black holes imply a more diluted dark matter production, which requires a larger mass $m_j$
to fit with the relic abundance constraints. The effect is the opposite if one decreases $q$ because, in this case, the memory burden effect is {\it delayed}, and a larger population of PBH decay
follows the semiclassical approximation as we can see in Fig.~\ref{fig:mj-min-f1} (right). In the limit $q\rightarrow 0$, we recover exactly the semiclassical limit. Moreover,
Note that once $\beta>\beta_{\rm c}$, the relic abundance no longer depends on $\beta$.

\begin{figure}
    \centering
    \includegraphics[scale=.57]{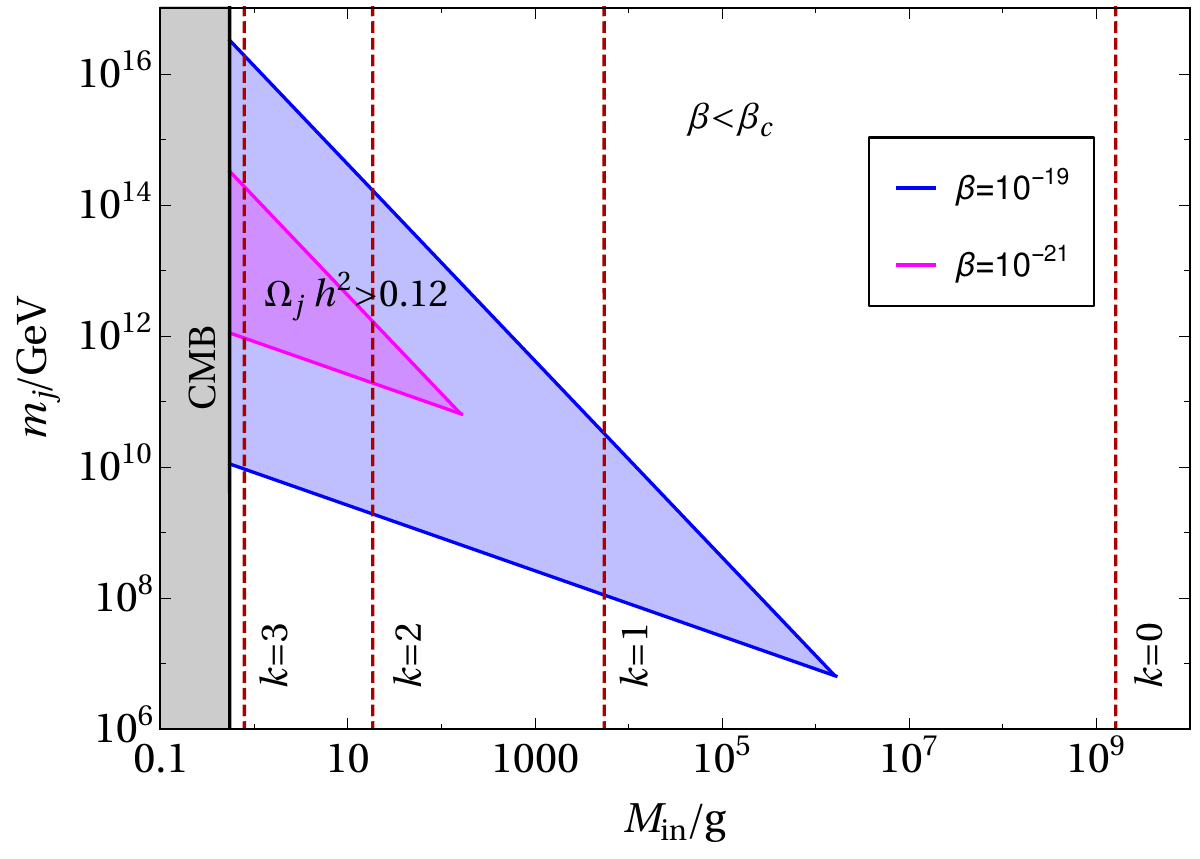}
    \caption{\it The values of the dark matter mass are plotted here as a function of the formation mass of PBHs for the case where the evaporation happens during radiation domination, i.e., $\beta<\beta_{\rm c}$. 
    The black lines represent the minimum mass possible for PBHs calculated considering the maximum energy scale of inflation.
   The vertical red dashed lines represent the maximum $\Min$ values allowed to be consistent with the BBN bound.
    The shaded regions correspond to dark matter overproduction, $\Omega_j\,h^2>0.12$.
    We have chosen $q=1/2$ and plotted two different values of $\beta=10^{-19}$ and $10^{-21}$, which are shown in blue and magenta, respectively.
    }
    \label{fig:mj-min-f2}
\end{figure}

\subsubsection{$\beta<\beta_{\rm c}$}

For $\beta<\beta_{\rm c}$, PBHs evaporate during radiation domination, so there is no PBH domination. 
The Hubble parameter at the point of evaporation is then given by
\bea
H^2(\aev)=\frac{\rho_{R}(\aev)}{3\Mpl^2}
=\frac{g_{\rm *} \pi^2 \Tev^4}{90 \Mpl^2}
=\frac{ \,\left({\gammabh^k}\right)^2}{4}\,,
\label{Eq: beta<betac}
\eea
which gives the evaporation temperature 
\bea 
\Tev= \Mpl\l(\f{3}{4\alpha}\r)^{1/4}
\l[(3+2k)2^k\epsilon\l(\f{\Mpl}{q\,\Min}\r)^{3+2k}\r]^{1/2}.
\label{eq:evap_temp_1}
\eea 
 During radiation dominated era, taking $H=1/(2\,t)$ and $a\propto t^{1/2}$, using
\beq
\nbh(\aev)=\nbh(\ain)\left(\frac{\ain}{\aev}\right)^3=\frac{\beta \rho_R^{\rm in}}{\Min}\left(\frac{\ain}{\aev}\right)^3
,~~~\frac{\ain}{\aev}=\sqrt{\frac{H(\aev)}{\hin}}=\sqrt{\frac{\gammabh^k}{2\hin}}\,,
 \eeq
 and $\hin$ given by Eq.~\eqref{Eq:min},
 we get the number density of the PBH to be
 \bea\label{Eq:numberblbc}
\nbh(\aev)= 3\,\beta\,\Mpl^3 
\l(\f{\pi\,\gamma}{2}\r)^{1/2}\l(\f{(3+2k)2^k\,\epsilon}{q^{3+2k}}\r)^{3/2}\l(\f{\Mpl}{\Min}\r)^{6+3k}.
 \eea 
 
Finally Combining Eq.~\eqref{Eq:omegah2} together with Eqs.~\eqref{eq:evap_temp_1} and \eqref{Eq:numberblbc}, we have DM abundance today for $m_j<\Tbh^{\rm in}$
 \begin{equation}
\f{\Omega_jh^2}{0.12}=2.54\times10^6\,\beta\,\xi\,g_j\, \l(\f{\Min}{\Mpl}\r)^{\frac{1}{2}} 
\,\frac{m_j}{\text{GeV}} 
\simeq \xi\,g_j\left(\frac{\beta}{10^{-20}}\right)\left(\frac{\Min}{1\,\rm{g}}\right)^\frac{1}{2}\left(\frac{m_j}{8.2\times 10^{10}\,\rm{GeV}}\right)
\,.
\label{Eq:omegahbl}  
\end{equation}
On the other hand, by doing the same analysis, one can arrive at the following expression for $m_j>T_{\rm BH}^{\rm in}$

\begin{equation}
\f{\Omega_jh^2}{0.12}= 1.47\times10^{43}\,{\beta\,\xi\,g_j} \l(\f{\Mpl}{\Min}\r)^{\frac{3}{2}} 
\,\frac{\text{GeV}}{m_j}
\simeq \xi\,g_j\left(\frac{\beta}{10^{-20}}\right)\left({\frac{1\,\rm{g}}{\Min}}\right)^\frac{3}{2}\left(\frac{1.3\times 10^{15}\,\rm{GeV}}{m_j}\right)
\,.
\label{Eq:omegahmjgPBH}  
\end{equation}
From the above expressions for two different cases Eqs.~\eqref{Eq:omegahbl} and \eqref{Eq:omegahmjgPBH}, it is interesting to note that the DM relic abundance is independent of the memory burden parameter $k$. This is easily understandable as, in this case, the total number of produced dark matter particles is the same irrespective of the physics behind the PBH decay, and further, the dilution effect is exclusively due to the radiation background.

We show in Fig.~\ref{fig:mj-min-f2} the relic abundance constraint for $q=1/2$ and two different values of $\beta=10^{-19}$
and $10^{-21}$, well below $\beta_{\rm c}$ as one can see from 
Fig.~\ref{fig:beta-crit}. We note that for the case $\beta < \beta_{\rm c}$ the parameter space follows a triangular exclusion zone 
in the center limited by the cases $m_j < \Tbh^{\rm in}$ and $m_j > \Tbh^{\rm in}$. However, we indeed see that such shape of the constraints region does not 
depend on the burden parameter $k$ anymore except for setting the maximum value $\Min$ represented by red dashed lines. As stated earlier, this is due to the fact that the total number of dark matter particles produced is the same for both semiclassical and quantum-corrected PBH evaporation processes in conjunction with the radiation background. 
On the other hand, as expected, there exists a strong dependence
of the relic density on $\beta$, for this parameter determines the relative amount of PBHs present in the thermal plasma, which will eventually decay 
(proportionally to $\beta$) into dark matter.

{ Note that the PBH evaporation produces DM with a significant initial momentum. A higher initial momentum indicates a large free
streaming length, which might erase small-scale structures.  In fact, the classical limit on warm dark matter ($m_j\geq 3$ keV) arising from Lyman-$\alpha$ restrictions can impose constraints on the mass of the DM, which needs to be reexamined in this context. 
In this context, the possibility of becoming warm dark matter was first
discussed in \cite{Lennon:2017tqq}.
Recently,
Ref. \cite{Barman:2024iht} showed how the constraints from the warm dark matter further restrict the DM mass in the case of
memory-burdened PBHs.}

%%%%%%%%%%%%%%%%%%%%%%%%%%%%%%%%%%%%%%%%%%%%%%%%%%%%%%
\subsection{Dark matter from the stable PBHs with Hawking evaporation ({\it phase-I}) before BBN \label{sec:dm-stable-evap}}

\iffalse
\begin{figure}
    \centering
    \includegraphics[scale=.57]{betac_fpbh.pdf}
    \caption{\it The critical value of $\beta$ for $\fpbh=1$ and for PBH domination are plotted here. The solid lines represent $\beta_{\rm c}$ corresponding to $\fpbh=1$, and the dashed lines represent $\beta_{\rm c}$ for PBH domination to happen before radiation matter equality. We have plotted three different values of $q$ where $q=0.1,\,0.5$, and $0.9$ are plotted in blue, red, and magenta, respectively. We see that for each case the total dark matter to be PBHs, it inevitably leads to PBH domination. 
    }
    \label{fig:betac_fpbh}
\end{figure}
\fi

Let us now look into the detail of the situation where standard Hawking evaporation takes place {\it before} BBN 
while the memory burden effect delays the complete evaporation, making the PBHs decay {\it after} 
the present time, rendering them stable on the cosmological scale. 
In this case, the total dark matter relic ($\Omega_{\rm DM}$) has two contributions: the dark matter as stable fundamental particles ($\Omega_j$) produced from PBH evaporation computed in the previous sections 
{\it plus} the stable PBHs $\Omega_{\rm PBH}$ that contribute
also to the dark component of the Universe :
$\Omega_{\rm DM}=\Omega_j+ \Omega_{\rm PBH}$.
%$\Omega_j$ is the contribution from the products of evaporation and $\Omega_{\rm PBH}$ represents the relic for stable PBHs that behave as DM. 
This is attributed to the allowed region
in white tagged as ``Stable PBHs" in Fig.~\ref{fig: Mpbh k}.

To compute the dark matter abundance
generated by the evaporation during {\it phase-I}, $\Omega_j$, we can use the expression for present-day DM relic due to evaporation, Eq.~\eqref{Eq:omegah2},
taken at $\tq$, the time duration of {\it phase-I} :

\bea
\Omega_jh^2= 1.6\times 10^8\,\frac{g_0}{g_{\rm q}}\frac{ n_{ j}(\aq)}{T^3(\aq)}\,\frac{m_j}{\text{GeV}}\,,
\label{Eq:omegahhalf}  
\eea
where $g_{\rm q}$ is the effective number of degrees of freedom for the entropy at the end of 
{\it phase-I}, which we assume has the same value\footnote{Generalization for any $g_{\rm q}$ is straightforward.} as $g_{\rm ev}$. 
The ratio $n_j(\aq)/T^3(\aq)$ is given by\footnote{We supposed here $\beta < \beta_c$ as when stable PBHs contribute to the entire dark matter, PBH domination starts roughly at the standard radiation-matter equality.}

\bea \label{Eq:ratiophase-I}
\f{n_j(\aq)}{T^3(\aq)}=
N_j \f{\nbh(\ain)}{\Tin^3}=
N_j\frac{\beta\alpha\,\Tin}{\Min}
\,,
\eea 
where $\Tin$ is the radiation temperature at the point of formation, and we supposed the Universe 
is radiation-dominated during the semiclassical 
phase ($\beta<\beta_{\rm c}$). Following Eq.~\eqref{Eq:min} we have

\begin{figure}
    \centering
    \includegraphics[scale=.37]{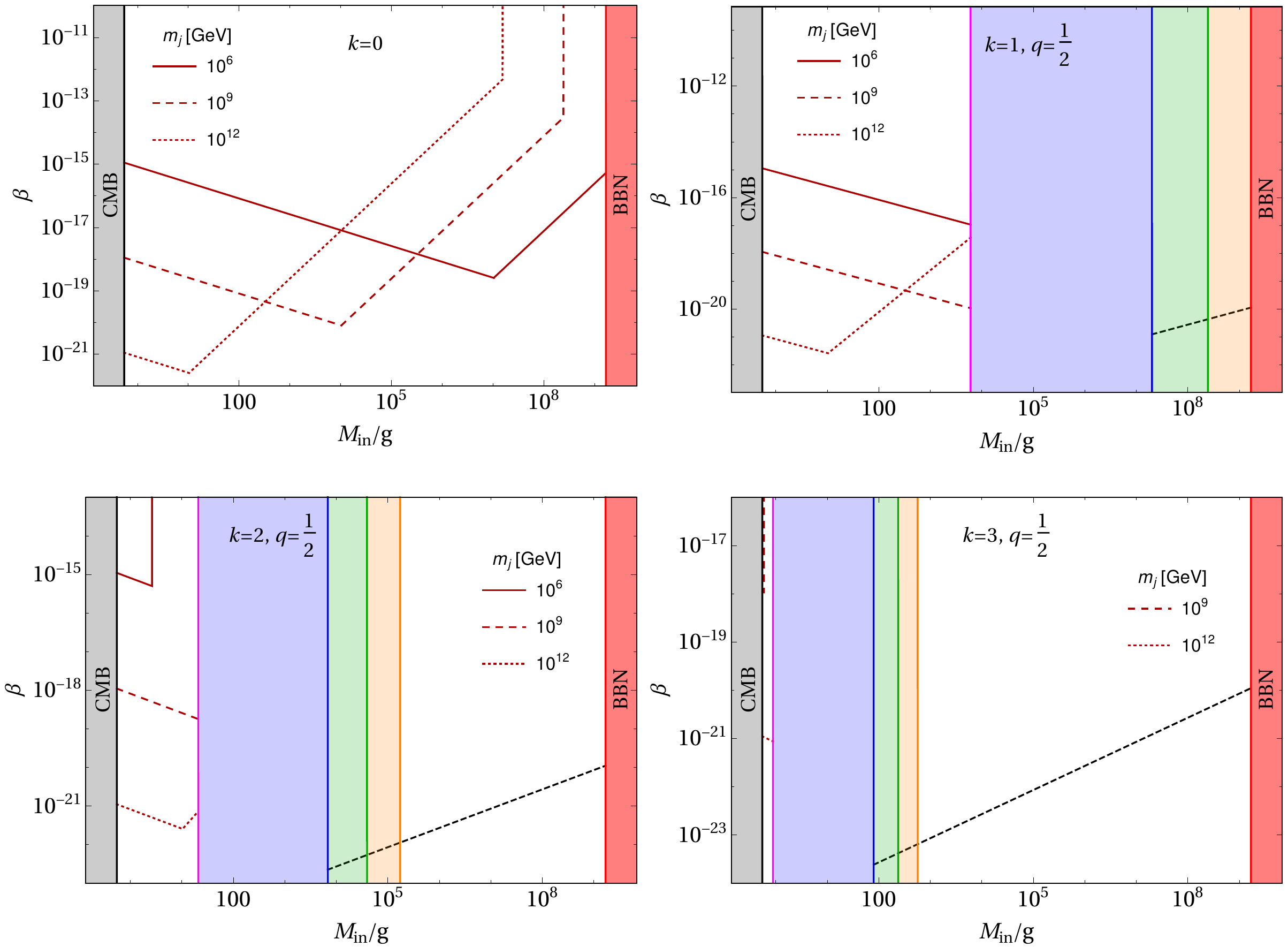}
    \caption{\it The critical values of $\beta$ corresponding to the total dark matter density are plotted in brown as a function of PBH mass when the dark matter is emitted from the evaporation of PBHs before BBN. We have chosen three different values of dark matter mass, where $m_j=10^6$ GeV, $10^9$ GeV, and $10^{12}$ GeV are plotted in solid, dashed, and dotted lines, respectively. The black dashed lines correspond to the critical $\beta$ when the stable PBHs contribute to the total dark matter energy density. The black-shaded region is excluded from the minimum PBH mass possible, corresponding to the highest energy scale of inflation. The red-shaded region indicates the PBH masses whose phase-I of evaporation ends after BBN. The blue regions correspond to the masses for which PBHs evaporate after BBN. The green and orange shaded regions are constraints coming from CMB and extragalactic $\gamma$-rays where $\fpbh<1$.
    }
    \label{fig:betac_fpbh}
    \end{figure}

\bea \label{Eq:Tin}
\Tin=\left(\frac{48\pi^2\gamma^2}{\alpha}\right)^{{1}/{4}}\,\sqrt{\frac{\Mpl}{\Min}}\,\Mpl
\,,
\eea
Now, upon substitution of Eqs.~\eqref{Eq:ratiophase-I} and \eqref{Eq:Tin} into \eqref{Eq:omegahhalf}, the final expression for DM relic from evaporation 
\bea \label{Eq:evap}
\f{\Omega_j\,h^2}{0.12}\simeq1.5\times 10^{9}N_j\,\beta \l(\f{\Mpl}{\Min}\r)^{3/2} \f{m_j}{\rm GeV}
\,,
\eea 
Where the total number of particles emitted per PBHs from {\it phase-I}, are given by Eqs.~\eqref{eq:N1j 1}
and \eqref{Eq:njphase1more}
\bea 
N_j= \f{15\,\xi\, g_j \zeta(3)}{g_{\ast}(\Tbh)\,\pi^4}
\begin{cases}
    (1-q^2)\f{\Min^2}{\Mpl^2}, & {\rm for}~m_j<\Tbh^{\rm in}\\
    \f{\Mpl^2}{m_j^2}-\f{q^2 \Min^2}{\Mpl^2}, &
    {\rm for}~    \f{\Tbh^{\rm in}}{q} > m_j>\Tbh^{\rm in}
\end{cases}.
\eea 

We then need to consider the contribution from stable PBHs. Their relic abundance can be written as

\bea\label{Eq:stablerelic}
\Omega_{\rm PBH}\,h^2=1.6\times 10^8\,\frac{g_0}{g_{\rm q}}\frac{ \rho_{\rm BH}(\aq)}{T^3(\aq)}\,\frac{1}{\text{GeV}}\,.
\eea
Connecting the end of {\it phase-I} with the formation point, one can find the ratio
\bea \label{Eq:stableBH}
\frac{ \rho_{\rm BH}(\aq)}{T^3(\aq)}=\frac{ q\,\rho_{\rm BH}(\ain)}{T^3(\ain)}=q\,\beta\,\alpha\,\Tin\,,
\eea
where we assumed that the relativistic degrees of freedom associated with radiation at the point of formation and at the end of the semiclassical phase are the same. Finally, combining Eq.~\eqref{Eq:Tin}
with Eq.~\eqref{Eq:stableBH}, and substituting into Eq.~\eqref{Eq:stablerelic}, we obtain the density of stable PBHs to be 
\bea \label{Eq:stableDM}
\f{\Omega_{\rm PBH}\,h^2}{0.12}= 3.5 \times 10^{27}\,q\,\beta 
\l(\f{\Mpl}{\Min}\r)^{1/2}
\simeq q \left(\frac{\beta}{1.4\times 10^{-25}}\right)
\left(\frac{1~{\rm g}}{\Min}\right)^{1/2}
\,.
\eea 
Note that the quantity on the left side is also described as $\fpbh=\Omega_{\rm PBH}\,h^2/0.12$, which is the fraction of total dark matter that comes from stable PBHs today. 
Finally, the total dark matter relic would be the sum of the contribution from evaporation in {\it phase-I}, Eq.~\eqref{Eq:evap} and from the sable PBHs which acts as dark matter, \eqref{Eq:stableDM}, which gives

\bea 
\f{\Omega_{\rm DM}\,h^2}{0.12}= 3.5 \times 10^{27}\,\beta 
\l(\f{\Mpl}{\Min}\r)^{1/2}
\l[q+N_j\f{m_j}{\Min}\r]\,,
\label{eq:tot_dm}
\eea 
which is the key equation of our paper. It summarizes the amount of dark components due to stable PBH {\it and} its dark products. We expect domination of the dark products for small $\Min$, whereas PBH relics 
dominates for larger $\Min$.

%%%%%%%%%%%%%%%%%%%%%%%%%%%%%%%%%%%%%%%%%%%%%%%%%%%%%%
\begin{figure}
    \centering
    \includegraphics[scale=.38]{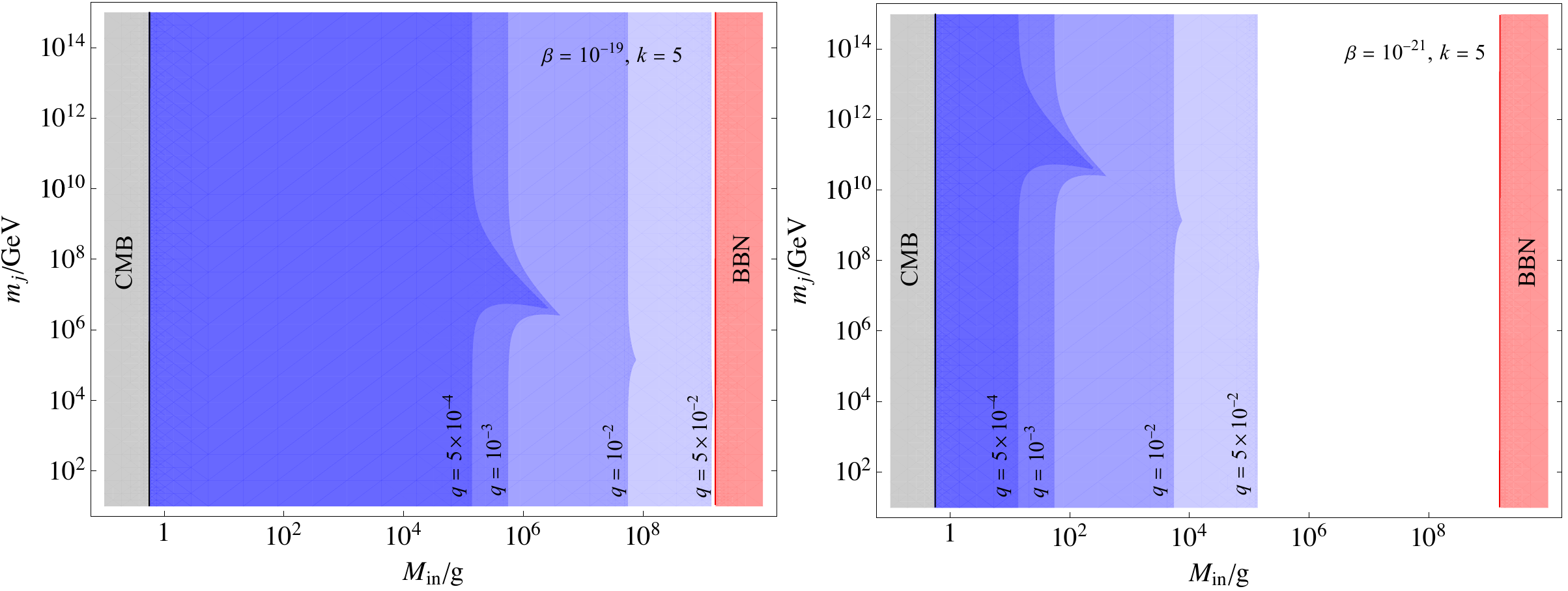}
    \caption{\it 
    Dark matter mass as a function of the PBH mass
    taking into account the contribution 
    from both the evaporation product and the stable PBHs.
    We have chosen two 
    different values of $\beta$ ($\beta=10^{-19}$-left and 
    $\beta=10^{-21}$-right). In both cases, we shaded the excluded (overdense) region for four different values of $q$
    ($q=5\times10^{-2},\,10^{-2}, \, 10^{-3}$ and $5\times 10^{-4}$).
    The grey-shaded region is
    excluded from the minimum PBH mass possible, corresponding to the highest energy scale of inflation.
    The red-shaded region indicates the PBH masses whose phase-I of evaporation ends after BBN.
    We see that for $q\gtrsim 10^{-2}$, the contributions to the total DM due to the evaporation are negligible compared to the contributions form the stable PBHs, see the text for detail. }
    \label{fig:betac_fpbh2}
\end{figure}
%%%%%%%%%%%%%%%%%%%%%%%%%%%%%%%%%%%%%%%%%%%%%%%%%

Indeed, to summarize, we show for comparison 
in Fig.~\ref{fig:betac_fpbh}
the parameter space allowed by the relic abundance constraint in the plane ($\Min$, $\beta$), 
in three cases : 
without taking into account the burden effect ($k=0$), and with burden effect for $k=1,2$ and $3$ for $q=1/2$
and $m_j=10^6$, $10^9$ and $10^{12}$ GeV.
We first distinguish the two white regions, for low $\Min$ and larger $\Min$, corresponding to the ``Evaporate before BBN" and ``Stable PBHs" of Fig.~\ref{fig: Mpbh k}. For $k=0$, in the mass range
$1~{\rm g} \lesssim \Min \lesssim 10^9~{\rm g}$
the PBHs decay completely before BBN, and the correct relic abundance is accomplished, depicted by the brown lines
by their decay product to a fundamental dark component in the first panel of Fig.~\ref{fig:betac_fpbh}.
They follow the behavior $\beta \propto \Min^{-1/2}$, see Eq.~\eqref{Eq:omegahbl} when 
$m_j < \Tbh^{\rm in}$ ($\Min < \frac{10^{13}{\rm GeV}}{m_j}$ g), and $\beta \propto \Min^{3/2}$, see Eq.~\eqref{Eq:omegahmjgPBH}
when $m_j >\Tbh^{\rm in}$ ($\Min > \frac{10^{13}{\rm GeV}}{m_j}$ g).

We recognize the same region for $k=1$ but with a restricted mass range, and the restriction is due to longer PBHs lifetime attributed to the memory burden effect, rendering  PBHs of mass $ 10^4 \,\mbox{g} < \Min < 10^7 \mbox{g}$ to decay after BBN and before the present time. In addition to that, the mass range from $ 10^7 \,\mbox{g} < \Min < 10^9\, \mbox{g}$ are stable PBHs with the first phase before BBN but still restricted by the constraints from the extragalactic $\gamma$ rays and CMB anisotropy shown in the orange and green shaded regions of Fig.~\ref{fig:betac_fpbh}. On the other hand, for $k=2$, a new unconstrained PBH mass window opens for large mass within $10^5 \mbox{g} < \Min < 10^9 \mbox{g}$, which survives at present due to the quantum effect and populates as total DM component,
while in the mass range $1\, \mbox{g} < \Min < 20\, \mbox{g}$, still keeping the possibility of the DM to be populated by the dark matter product from PBH decay.
In the case of stable PBH, the relic abundance is given by $\beta \propto \Min^{1/2}$ as one can see from Eq.~\eqref{Eq:stableDM}. Finally, only PBH as a dark matter is allowed for $k\geq 3$, with PBH relic densities not depending on $k$. This set of 4 figures nicely summarizes our work. Note that we set the $q$ value at $1/2$ to do this analysis.

However, to complete our analysis, we also show the dependence on
$q$ in Fig.~\ref{fig:betac_fpbh2}, for $\beta=10^{-19}$ (left) and 
$\beta=10^{-21}$ (right), corresponding to the values used in Fig.~\ref{fig:mj-min-f2} for comparison. The shadowed region is excluded because it corresponds to parameter space, where there is an overproduction of dark matter. The limit $q=0$ on the left region corresponds to the semiclassical approximation, i.e., the memory burden effect never occurs. This limit corresponds to 
the situation where $\sim 100\%$ of the dark matter is composed of the decay products of PBH because there are no stable PBHs in this mass range.
We can recognize the peak of the triangle shape of Fig.~\ref{fig:mj-min-f2} 
appearing at $\Min\simeq 10^6$ g 
($\simeq 100$ g) for $\beta=10^{-19}$ ($\beta=10^{-21}$).
For larger value of $q$, we know from Fig.~\ref{fig:betac_fpbh}
that for $\beta=10^{-19}$, PBH cannot be a dark matter candidate, as they would overpopulate the Universe. That explains why all the regions are shadowed in Fig.~\ref{fig:betac_fpbh2} left. However, for $\beta=10^{-21}$, we can 
evaluate form Fig.~\ref{fig:betac_fpbh} bottom--left that $\Min$ 
should be larger than $\sim 10^7$ g for $q=1/2$. But we see from Fig.~\ref{fig:betac_fpbh2} right that a larger region is allowed, up to the range $10~{\rm g}\lesssim \Min \lesssim 10^9~{\rm g}$
for $q=5 \times 10^{-4}$. Indeed, for a fixed $\beta$ value, smaller values of $q$ induce a latter burden effect, and a smaller stable PBH mass is required to satisfy the total DM relic. In summary, for stable PBHs with $q$ values $q\leq 10^{-2}$, there is a possibility that DM from PBH decay also contributes to the total DM relic, depending on the mass of the DM. However, for higher $q$ values $q\geq 10^{-2}$, the dark matter from stable PBHs always dominates over the decay products.

%%%%%%%%%%%%%%%%%%%%%%%%%%%%%%%%%%%%%%%%%%%%%%%%%%%%%%%%%%%%%%%%%%%%%%%%%%%%%%%
\section{Conclusions}
\label{Sec:conclusion}

We computed the relic abundance in the presence of PBH beyond the semiclassical approximation. Indeed, when the mass of a BH reaches a certain fraction $q$ of its initial value, the backreaction can not 
be ignored, and it can potentially reduce the evaporation rate by the inverse power law of its entropy $S^{-k}$.
Taking into account this effect due to the memory burden effect, 
we added the dark component produced by the decay of PBHs during the early stage to the contribution of stable PBHs whose lifetime 
is extended due to the quantum corrections. Since the lifetime of the BH is extended depending on the memory burden parameters $k$ 
and $q$, for the scenario of PBH domination, the duration of the PBH domination is also extended. As a result, the allowed range for DM mass $m_j$ is extended as we increase $k$ for a fixed $q$ or increase  $q$ for a fixed $k$; that is also reflected in Fig. \ref{fig:mj-min-f1}. Whereas, 
in the case of PBH decay in a background dominated by radiation, the memory burden parameters limit the highest permitted BH mass 
that completes the decay process before BBN and has no effect on the dark matter mass range, which we can see in Fig. \ref{fig:mj-min-f2}. Depending on the strength of the burden effect, we show 
that PBH decay products can satisfy the relic density constraints for PBH masses below $\lesssim (6\times 10^3,\,20,\,1)$ g for $k=(1,\,2,\,3)$, respectively. Moreover, with an increase in $k > 3$, there is no DM parameter space from PBH decay. Note that all of these mentioned values are quoted for considering $phase-I$ 
being complete after half-life, $q=1/2$. On the other hand, PBH mass in the range $10^3-10^9$ g can be sufficiently stable and 
numerous to fulfill the relic abundance of DM for $k=3$. The range of allowed masses even extends to $1~{\rm g}-10^9~{\rm g}$ for $k\geq5$, see Fig.~\ref{fig: Mpbh k}. 
We also show that decreasing the value of $q$ down to $\sim10^{-2}$
reopens the dark matter parameter space from PBH decay, then we need to consider the contribution from both PBH decay and stable PBHs. 
Our results are thus nicely summarized in Figs. \ref{fig: Mpbh k}, \ref{fig:betac_fpbh} and \ref{fig:betac_fpbh2}, with Eq.~\eqref{eq:tot_dm}.

%%%%%%%%%%%%%%%%%%%%%%%%%%%%%%%%%%%%%%%%%%%%%%%%%%%%%%%%%%%%%%%%%%%%%%%%%%%%%%%
\section*{Acknowledgments}
%%%%%%%%%%%%%%%%%%%%%%
The authors want to warmly thank Donald Kpatcha, Simon Clery, Mathieu Gross, and Lucien Heurtier for extremely useful discussions during the Astro@Paris-Saclay symposium 2023.
This project has received support from the European Union's Horizon 2020 research and innovation program under the Marie Sklodowska-Curie grant agreement No 860881-HIDDeN and the CNRS-IRP project UCMN. 
MRH wishes to acknowledge support from the Science and Engineering Research Board (SERB), Government of India (GoI), for the SERB National Post-Doctoral fellowship, File Number: PDF/2022/002988. 
SM wishes to thank the Indian Institute of Technology (IIT) Madras, Chennai, India, for support through the Exploratory Research Project RF22230527PHRFER008479. 
SM also wishes to acknowledge the support from project SB22231259PHETWO008479 for the travel grand for Astro@ParisSaclay symposium 2023.
DM wishes to acknowledge support from the Science and Engineering Research Board
(SERB), Department of Science and Technology (DST), Government of India (GoI), through the Core Research Grant CRG/2020/003664.

%%%%%%%%%%%%%%%%%%%%%%%%%%%%%%%%%%%%%%%%%%%%%%%%%%%%%%%%%%%%%%%%%%%%%%%%%%%%%%%
%%%%%%%%%%%%%%%%%%%%%%%%%%%%%%%%%%%%%%%%%%%%%%%%%%%%%%%%%%%%%%%%%%%%%%%%%%%%%%%

\bibliographystyle{JHEP}
\bibliography{references}

%%%%%%%%%%%%%%%%%%%%%%%%%%%%%%%%%%%%%%%%%%%%%%%%%%%%%%%%%%%%%%%%%%%%%%%%%%%%%%%
\end{document}